\documentclass[a4paper,10pt]{article}

\usepackage{hyperref}

\usepackage{amssymb,amsmath,amsfonts,amsthm,a4wide}

\usepackage{epsfig,psfrag}

\newtheorem{theorem}{Theorem}

\newtheorem{problem}{Problem}

\newcommand{\RR}{\mathbb{R}}

\newcommand{\dd}{\mathrm{d}}
\newcommand{\DD}{\triangle}
 \,
 \, 
\newcommand{\ie}{i.\,e.}
\newcommand{\eg}{e.\,g.}

\numberwithin{equation}{section}

\begin{document}

\title{{\bf Mean-Field Limit of Quantum Bose Gases and Nonlinear Hartree Equation\footnote{Notes to a lecture presented by J.~F.~in the PDE seminar at X-Palaiseau, Paris, April 2004.}}}
\author{{\sc {\large  J\"urg Fr\"ohlich and Enno Lenzmann}}}
\maketitle

\begin{abstract}
We discuss the Hartree equation arising in the mean-field limit of large systems of bosons and explain its importance within the class of nonlinear Schr\"odinger equations. Of special interest to us is the Hartree equation with focusing nonlinearity (attractive two-body interactions). Rigorous results for the Hartree equation are presented concerning: 1) its derivation from the quantum theory of large systems of bosons, 2) existence and stability of Hartree solitons, and 3) its point-particle (Newtonian) limit. Some open problems are described.  
\end{abstract}

\newpage

\tableofcontents

\newpage

\section{Introduction}

We present an overview of rigorous results and open problems concerning the {\em Hartree equation} 
\begin{equation*} \tag{HE} \label{eq-HE}
 i \partial_t \psi = - \DD \psi + V \psi + \big ( \Phi \ast | \psi |^2 \big ) \psi \, , \qquad x \in \RR^3 \, ,
\end{equation*}
where $\psi = \psi(t,x)$ is complex-valued, and $V$ and $\Phi$ are real-valued functions on $\RR^3$. Furthermore, $\DD$ denotes the Laplacian and $\ast$ convolution of functions on $\RR^3$. 

Equation (\ref{eq-HE}) has many interesting applications in the quantum theory of large systems of non-relativistic bosonic atoms and molecules. In fact, this equation arises in the study of the {\em mean-field limit} of such systems, \ie, of a regime where the number of bosons is very large, but the interactions between them are weak. It is therefore of considerable interest to extend mathematical methods originally developed for nonlinear Schr\"odinger equations with local nonlinearities to the study of equation (\ref{eq-HE}) and to develop new methods.

More specifically, the work reviewed in these notes has been motivated by the following physical questions.

\begin{itemize}
\item[(A)] What is the behavior of many bosons with attractive two-body interactions in the mean-field limit?
\item[(B)] How does quantum mechanics reproduce Newtonian mechanics of point-particles in limiting regimes?
\end{itemize}

In these notes we review results providing partial answers to these questions. Our discussion is arranged as follows.

\begin{itemize}

\item In Section \ref{sec-mf}, we discuss the {\em mean-field limit} of a bosonic system and the emergence of the Hartree equation (\ref{eq-HE}) in this limiting regime. We present rigorous results, physical applications, and two open problems related to this limit.

\item Section \ref{sec-hamilton} contains a short review of the Hamiltonian structure connected with (\ref{eq-HE}). We also review some conservation laws and various identities.

\item In Section \ref{sec-solitons}, we study solitary wave solutions of equation (\ref{eq-HE}), which we call {\em ``Hartree solitons"}. The main mathematical results are summarized in Theorems \ref{th-existence} and \ref{th-stability} asserting existence and orbital stability of Hartree ground state solitons for attractive, ``subcritical" potentials $\Phi$. These results are proven by using  variational methods. Uniqueness of Hartree ground states in general situations is still open and posed as a problem. At the end of this section, we address the delicate issue of {\em asymptotic stability} and suggest an open problem.  

\item In Section \ref{sec-pplimit}, we discuss the {\em point-particle (Newtonian) limit} of (\ref{eq-HE}). That is, by considering a suitable limiting process, we derive Newton's equations of motion for a system of point-like particles corrected by a term representing ``radiation damping". 

\item In Section \ref{sec-end}, we draw attention to some further open problems, and offer some conclusions and an outlook.
\end{itemize} 

\noindent
{\em Notation and Units.}
In what follows, physical space is chosen to be $\RR^3$, but the discussion can be extended to $\RR^d$, for $d \geq 3$. Throughout this text, Lebesgue spaces of complex-valued functions on $\RR^3$ are denoted by $L^p(\RR^3)$, with norm $\| \cdot \|_p$, $1 \leq p \leq \infty$. The Sobolev space $H^1(\RR^3) := \left \{ \psi \in L^2(\RR^3) : \| \nabla \psi \|_2 < \infty \right \}$ plays an important r\^ole. Physical units are chosen such that Planck's constant $\hbar = 1$.\\

\noindent
{\em Acknowledgments.} J.~F.~thanks all his collaborators in work on nonlinear dynamics, in particular on nonlinear Schr\"odinger and Hartree equations, and especially I.~M.~Sigal and T.~Spencer, for many stimulating and fruitful discussions. He thanks the organizers of the PDE seminar at Ecole Polytechnique for the opportunity to present a summary of that work. 

E.~L.~thanks J.~Shatah for introducing him to nonlinear Schr\"odinger equations in an inspiring way.

\section{Mean-Field Limit} \label{sec-mf}

We consider a system of $N$ spinless, identical, non-relativistic bosons with two-body interactions given by a potential, $\kappa \Phi(x)$, where $\kappa \geq 0$ is a coupling constant. These bosons are also subject to a time-independent, external potential, $\lambda V(x)$, with coupling constant $\lambda \geq 0$. If the mutual interactions among these bosons are weak, in a sense to be made precise below, the system can be interpreted as a {\em quantum Bose gas} under the influence of an external potential called ``trap" in condensation experiments. 

For simplicity, we only consider pure states of this $N$-particle system. The dynamical evolution of a pure state, $\Psi_N$, is governed by the linear Schr\"odinger equation
\begin{equation} \label{eq-SE}
i \partial_t \Psi_N = H_N \Psi_N \, , 
\end{equation}
where $\Psi_N = \Psi_N(t,x_1,\ldots,x_n)$, with $x_j \in \RR^3$, $1 \leq j \leq N$. The $N$-particle Hamiltonian is given by
\begin{equation} \label{eq-HN}
H_N := \sum_{j=1}^N  \big ( - \! \frac{1}{2m} \DD_{x_j} + \lambda V(x_j) \big ) + \kappa \sum_{ i < j }^N \Phi(x_i-x_j) \, , \qquad \mbox{on} \quad L^2(\RR^3)^{\otimes_s N} \, ,
\end{equation} 
where $\otimes_s$ denotes the symmetric tensor product. In equation (\ref{eq-HN}), $\DD_{x_j}$ denotes the Laplacian associated with the $j$th copy of $\RR^3$; the boson mass is $m$.

Since the total energy of the system scales like $\mathcal{O}(N)+ \kappa \mathcal{O}(N^2)$, the energy per particle is $\mathcal{O}(1)$ if the coupling obeys $\kappa = \mathcal{O}(N^{-1})$. Thus, one defines the {\em mean-field limit}  to correspond to
\begin{equation} \boxed{
N \rightarrow \infty \quad \mbox{and} \quad \kappa \rightarrow 0 \, , \qquad \mbox{such that} \quad \nu := \kappa \cdot N = \mathrm{const.} }
\end{equation}
As shown below, this limit leads to the Hartree equation as an effective description of the dynamics of so-called {\em coherent states}, or {\em condensates}, of the system.

\noindent
{\em Remarks.} 1) Note that the statistics of bosons dictates this scaling, whereas a fermionic system would require a different scaling $\kappa = \mathcal{O}(N^{-1/3})$.

2)  There is a second limiting regime referred to as {\em ``classical limit"}. It is constructed as follows. Approach the mean-field limit, rescale time according to $t = \sqrt{m} \tau$, and take $m  \rightarrow \infty$; (this is equivalent to $\hbar \rightarrow 0$). In this limit, the {\em Vlasov equation} emerges from (\ref{eq-SE}, \ref{eq-HN}); see \cite{graffi-03} and references given there.

\subsection{Heuristic Discussion} \label{subsec-heuristics}

We now demonstrate heuristically the emergence of the Hartree equation in the description of a bosonic system in its mean-field limit. For this purpose, we investigate the limiting regime of a {\em condensate}  where all -- except for $o(N)$ -- bosons are in the {\em same} one-particle state, which is described by a wave function $\psi \in L^2(\RR^3)$.  We choose a {\em coherent state} as an initial datum for (\ref{eq-SE}). That is, we pick $\psi_0 \in L^2(\RR^3)$, with $\| \psi_0 \|_2^2 = 1$, and introduce the $N$-particle state
\begin{equation} \label{eq-soc}
\Psi_N(t=0,x_1,\ldots,x_N) = \prod_{j=1}^N \psi_0(x_j) \, ,
\end{equation}
which apparently belongs to $L^2(\RR^3)^{\otimes_s N}$. Assuming that the linear Schr\"odinger equation (\ref{eq-SE}) approximately preserves the coherent nature of $\Psi_N(t)$ when $N$ becomes large, we can write
\begin{equation} \label{eq-psiN}
\Psi_N(t,x_1,\ldots,x_n) \simeq \prod_{j=1}^N \psi(t,x_j) \,  .
\end{equation}
Physically speaking, this means that correlation effects remain small. 

Approaching the mean-field limit, we expect that the potential per particle is given in terms of an effective mean-field potential, $v_\mathrm{eff}$, as follows
\begin{equation} \begin{split}
 v_\mathrm{eff}(t,x) & := \lambda V(x) + \frac{\nu}{N} \sum_{j=1}^N \int_{\RR^3} \Phi(x-x_j) | \psi(t,x_j)|^2 \, \dd x_j \\ & = \lambda V(x) + \nu \big ( \Phi \ast |\psi(t)|^2 \big )(x)  \, ,
 \end{split}
\end{equation}
where
\begin{equation}
(f \ast g)(x) := \int_{\RR^3} f(x-y) g(y) \, \dd y \, .
\end{equation}
From this heuristic discussion we conclude that the dynamical evolution of the bosonic system in its mean-field regime is described by the Schr\"odinger equation for the one-particle wave function, $\psi(t,x)$, with a potential term given by $v_{\mathrm{eff}}(t,x)$. Thus, we are led to the (nonlinear) {\em Hartree equation} as an effective description of the limiting dynamics:
\begin{equation*} \tag{HE}   \label{eq-HE2}
\boxed{  \begin{array}{c} \displaystyle i \partial_t \psi = -\DD \psi + \lambda V \psi + \nu \big ( \Phi \ast |\psi|^2 \big ) \psi \, , \\ 
\psi(0,x) = \psi_0(x) \,  , \end{array} }
\end{equation*} 
and we will choose units such that
\begin{equation} 
 m = 1/2 \, ,
 \end{equation}
 from now on. In (\ref{eq-HE}), the complex-valued function, $\psi=\psi(t,x)$, with  $x \in \RR^3$, describes the one-particle state of the condensate with initial state $\psi(0,x) = \psi_0(x)$. For convenience, we keep the coupling constants, $\lambda \geq 0$ and $\nu \geq 0$, explicit throughout the following. 

\subsection{Rigorous Results} \label{subsec-rigorous}

A derivation of the Hartree equation in the mean-field limit of many bosons has been achieved with mathematical rigor. The most delicate step to be taken in this analysis is to show that correlation effects actually do remain small for {\em all} times, \ie, approximation (\ref{eq-psiN}) is justified.

The first rigorous derivation of (\ref{eq-HE}) dates back to K.~Hepp's work \cite{hepp-74}, where  $\Phi$ is assumed to be smooth and bounded. A generalization valid for all bounded potentials and one-particle density matrices (thus admitting mixed states) is due to H.~Spohn \cite{spohn-80}. Recent work by L.~Erd\"os and H.-T.~Yau \cite{ey-01} covers a rigorous derivation for Coulomb potentials $\Phi(x) = \pm |x|^{-1}$. New simple proofs will appear in \cite{schwarz-04}.

We briefly sketch some of the main ideas in the rigorous works cited above. For this purpose, we introduce $m$-particle operators, $\mathcal{A}_m$, defined in terms of integral kernels, $a(x_1,\ldots,x_m; y_1, \ldots, y_m)$, which are assumed to be smooth, bounded functions on $\RR^{3m} \times \RR^{3m}$, $1 \leq m \leq N$. The expectation value of $\mathcal{A}_m$ in an $N$-particle state, $\Psi_N$, is given by
\begin{equation} \begin{split}
\langle \Psi_N, \mathcal{A}_m \Psi_N \rangle =  \int_{\RR^{3N} \times \RR^{3N}} & \overline{{\Psi}_N(x_1,\ldots,x_N)} \, a(x_1, \ldots, x_m; y_1,\ldots,y_m)   \\  & \Psi_N(y_1,\ldots,y_N) \prod_{j = m+1}^N \delta(x_j - y_j) \, \dd x \, \dd y \, .
\end{split}
\end{equation}
If the $N$-particle initial state is {\em coherent}, \ie,
\begin{equation} \label{eq-CS}
\Psi_N(t=0,x_1,\ldots,x_N) = \prod_{j=1}^N \psi_0(x_j) \, ,
\end{equation}
for some $\psi_0 \in L^2(\RR^3)$, then, under certain assumptions on $\Phi$, $V$ and $\psi_0$, 
\begin{equation} \label{eq-AExp} \begin{split}
\lim_{{N \rightarrow \infty \atop \nu = \kappa \cdot N = \mathrm{const.}}} \langle \Psi_N(t), \mathcal{A}_m \Psi_N(t) \rangle = \int_{\RR^{3m} \times \RR^{3m}} & a(x_1,\ldots,x_m;y_1,\ldots,y_m)  \\ & \prod_{j=1}^m \overline{\psi(t,x_j)} \, \psi(t,y_j) \, \dd x \, \dd y \, , \end{split}
\end{equation} 
as proven in the papers quoted above. Here $\Psi_N(t)$ denotes the solution of the linear Schr\"odinger equation (\ref{eq-SE}) with initial condition (\ref{eq-CS}), whereas $\psi=\psi(t,x)$ solves the Hartree equation (\ref{eq-HE}) with initial datum $\psi_0$. Equation (\ref{eq-AExp}) justifies our heuristic discussion of Section \ref{subsec-heuristics}: All relevant information on the dynamics of coherent states of the bosonic system in its mean-field limit can be obtained from the Hartree equation.

For our mathematical analysis of the Hartree equation it is convenient to introduce a class of potentials defined as follows. A potential $\Phi$ is called {\em subcritical} if
\begin{equation} \label{eq-sub}
 \Phi \in L^p(\RR^3) + L^\infty(\RR^3) \, , \qquad \mbox{for some} \quad  p > 3/2 \, ,
\end{equation} 
and likewise for $V$. Furthermore, we define the {\em energy space} for (\ref{eq-HE}) to be 
\begin{equation} \label{eq-X}
 X := \left \{ \psi \in H^1(\RR^3) :  V |\psi|^2 \in L^1(\RR^3) \right  \} \, , 
\end{equation}
equipped with the norm $\| \psi \|_X := \| \psi \|_{2} +\| \nabla \psi \|_2 + \| |V|^{1/2} \psi \|_2$.

Referring to \cite{cazenave-03} and references therein, we can state the following result concerning the Cauchy problem for (\ref{eq-HE}) in energy space.

\begin{theorem} \label{th-gwp} {\bf [Global Well-Posedness]} Let $\Phi$ be real-valued, radial, and subcritical. Assume that $V$ is real-valued and satisfies one of the following assumptions.
\begin{enumerate}
 \item[a)] $V$ is subcritical (\ie, $V$ is a ``localized trap", or $V \equiv 0$).
 \item[b)] $V \in C^\infty(\RR^3)$, $V \geq 0$, and $|D^\alpha V| \in L^\infty(\RR^3)$ for all $|\alpha| \geq 2$ (\ie, $V$ is a smooth ``confining trap" with at most quadratic growth).
 \end{enumerate}
Then (\ref{eq-HE}) is globally well-posed in $X$. That is, for every initial datum $\psi_0 \in X$, there exists a unique solution $\psi \in C(\RR, X)$ depending continuously on $\psi_0$.
\end{theorem}

\noindent
 {\em Remarks.} 1) Note that if condition a) holds, then we have that $X \equiv H^1(\RR^3)$ by virtue of Sobolev's inequalities.
 
2) The study of the Cauchy problem for (\ref{eq-HE}) dates back to the work of J.~Ginibre and G.~Velo \cite{ginibre-80}.  

3) We expect that global well-posedness for appropriately chosen initial conditions can also be proven for external potentials growing more rapidly than $|x|^2$, as $|x| \rightarrow \infty$; cf.~Theorem \ref{th-existence} below.

4) The conditions of Theorem \ref{th-gwp} also imply that the $N$-particle Hamiltonian, $H_N$, is bounded from below due to the Kato-Sobolev inequalities. In addition, $H_N$ then defines a unique selfadjoint operator via its form sum; see, \eg, \cite{cfks-87}. \\

In the following discussion, we mainly consider attractive, subcritical $\Phi$'s. We derive existence of Hartree ground states and establish their orbital stability for such two-body potentials; see Theorems \ref{th-existence} and \ref{th-stability} below. We will, however, also comment on the {\em critical} case, $\Phi(x) =-|x|^{-2}$, which misses to be subcritical and leads to blow-up phenomena in (\ref{eq-HE}) of interest in physics.

\subsection{Physical Applications}

Here we describe three physical examples where the mean-field limit yields an appropriate description.

\subsubsection*{Repulsive Interactions} 
Bose-Einstein condensation (BEC) in gases with very weak {\em repulsive} two-body interactions can be found in systems of ${}^{87}$Rb or ${}^{23}$Na atoms. In this repulsive case, the approximation of $\Phi(x)$ by the delta function, $+\delta(x)$, is adequate, and (\ref{eq-HE}) becomes a cubic nonlinear Schr\"odinger equation of local type, \ie,
\begin{equation}
i \partial_t \psi = -\DD \psi + V \psi + |\psi|^2 \psi \, ,
\end{equation}
which is often referred to as the {\em Gross-Pitaevskii equation}; see \cite{lrsy-03} for a comprehensive review of rigorous results on BEC with repulsive interactions.\\

\subsubsection*{Attractive Interactions} 

BEC in gases with very weak {\em attractive} two-body interactions is observed, for example, in systems of ${}^7$Li atoms, as long as the gas in the trap has a sufficiently low density. But even for $\lambda = 0$ there is a critical number, $\nu_* \geq 0$, such that if
\begin{equation} \label{eq-kN}
  \kappa N > \nu_* 
\end{equation}
bound clusters of atoms exist. This feature is reflected in the existence of Hartree solitons when no external potential is present; see Theorem \ref{th-existence}, below, for a precise statement. 

In experiments, a collapse process is observed if the number of ${}^7$Li atoms in the trap exceeds some threshold value. But we remark that, in a theoretical description of such collapse processes, it is no longer justified to treat bosonic atoms as point-particles. Their internal structure, ionization processes and recombination, and interactions with the electromagnetic field are important when the density of the system becomes large. Neither the Gross-Pitaevskii equation (with focusing nonlinearity) nor the Hartree equation take such processes into account. One may expect that, at best, the Gross-Pitaveskii equation (with focusing nonlinearity) and the Hartree equation (with $\Phi =$ an attractive van der Waals potential) give a qualitatively satisfactory account of the {\em onset} of the collapse process for $\kappa N > \nu_*$.\\

\subsubsection*{Boson Stars} 

Boson stars are objects, at least theoretical ones, made of bosons subject to their own gravity. It is reasonable to use mean-field theory to describe such objects. A model of a boson star in its non-relativistic regime is given by 
\begin{equation} \label{eq-bsnr}
  i \partial_t \psi = -\DD \psi - \nu \big ( |x|^{-1} \ast |\psi|^2 \big ) \psi \, ,
\end{equation}
which is also referred to as the {\em Schr\"odinger-Newton equation} (or the {\em Choquard equation} in another physical context; see \cite{lieb-77}). Existence and stability of Hartree solitons arising from ground states can be shown for (\ref{eq-bsnr}); see Theorems \ref{th-existence} and \ref{th-stability} below. We refer to \cite{harrison-03} for a numerical study of (\ref{eq-bsnr}).

An improved model of a boson star arises when high velocities of the bosons are taken into account by incorporating special relativistic effects, but still assuming Newtonian gravity. That is, we consider the semi-relativistic Hartree equation
\begin{equation} \label{eq-bs}
 i \partial_t \psi = \big ( \sqrt{-\DD + m^2} - m \big ) \psi - \nu \big ( |x|^{-1} \ast |\psi|^2  \big ) \psi \, ,
\end{equation}
with $H^{1/2}(\RR^3)$ as energy space. Equation (\ref{eq-bs}) leads to a {\em Chandrasekhar type} theory of boson stars. See \cite{liebyau-87} for a study of the time-independent equation; an extended discussion of (\ref{eq-bs}) will appear in \cite{lenzmann-bs-04}. Note that, since the relativistic kinetic energy has a dimension of length$^{-1}$, the two-body potential, $\Phi(x)=-|x|^{-1}$, is critical, large initial data can lead to blow-up phenomena, which can be thought of as an indication of gravitational collapse. Of course, this model has its limitations. Nevertheless, it is useful to explore the onset of collapse and structure formation of bosonic matter by studying equations such as (\ref{eq-bs}).

\subsection{Two Open Problems about the Mean-Field Limit}

\subsubsection*{Mean-Field Spectral Analysis} 

As mentioned above, the Hartree equation captures the dynamics of a bosonic system in the mean-field limit. We now ask how the bound state energies of the $N$-particle Hamiltonian, $H_N$, can be derived from an effective one-particle Schr\"odinger operator, as the mean-field limit is approached. 

To gain some insight into this problem, we propose the following simplified ``effective" description. We define the ground state energy of $H_N$ by
\begin{equation}
E^0_N := \inf \sigma(H_N) \, .
\end{equation}
This quantity can be related to the Hartree equation (\ref{eq-HE}) as follows. On the basis of results in \cite{liebyau-87}, we expect that 
\begin{equation} \label{eq-E0N}
E^0_N = N e_0 + o(N) \, ,
\end{equation}
as $N \rightarrow \infty$, with $\nu = \kappa \cdot N$ fixed, where
\begin{equation} \label{eq-e0}
e_0 := \inf \left \{ 2 \mathcal{H}[\psi]  : \psi \in X , \, \| \psi \|_2^2 = 1 \right \} \, ,
\end{equation}
and $\mathcal{H}[\psi]$ denotes the Hartree Hamilton functional; see (\ref{eq-H}) below. 

As shown in Theorem \ref{th-existence} below, there exists a minimizer, $\psi = Q$, for problem (\ref{eq-e0}), which is called a {\em Hartree ground state} and satisfies the Euler-Lagrange equation
\begin{equation}
 -\DD Q + \lambda V Q +  \nu \big ( \Phi \ast |Q|^2 \big ) Q = \varepsilon_0 Q \,  .
\end{equation} 

Next, we propose to generalize equations (\ref{eq-E0N}, \ref{eq-e0}) to a relation between the bound state energies of $H_N$ and the spectrum of an effective one-particle Hamiltonian given by
\begin{equation}
H_{\mathrm{eff}} := -\DD + \lambda V + \nu \big ( \Phi \ast |Q|^2 \big ) \, .
\end{equation}
For simplicity, we assume that $H_{\mathrm{eff}}$ and $H_N$ have finitely many bound states (after splitting off the center-of-mass motion of $H_N$ if $\lambda = 0$). We consider the eigenvalue problem 
\begin{equation}
H_{\mathrm{eff}} \phi_i = \varepsilon_i \phi_i \, ,
\end{equation}
with $\{ \varepsilon_0, \varepsilon_1, \varepsilon_2, \ldots \}$ the eigenvalues of $H_{\mathrm{eff}}$ and $\{ \phi_0 = Q, \phi_1, \phi_2, \ldots \}$ the corresponding eigenvectors. The mean-field picture leads to the following conjecture: Every bound state energy of $H_N$ below its continuum threshold, $\Sigma_N$, is given by the ground state energy of $H_{N-j}$ plus a suitable combination of $j$ energies from the set of $\varepsilon_i$'s. More precisely, we expect that the discrete spectrum of $H_N$ is given by
\begin{equation} \label{eq-problem1}
\sigma_{\mathrm{d}}(H_N) = \left \{ E : E= E_{N-j}^0  + \sum_{\ell=1}^j  ( \varepsilon_{i_\ell} -\varepsilon_0 + e_0 )  + o(1) < \Sigma_N \, , \; 0 \leq j < N \right \} \, ,
\end{equation}
with $\Sigma_N = E^0_{N-1}$, as $N \rightarrow \infty$, with $\nu = \kappa \cdot N$ fixed. Thus, we propose the following problem.
\begin{problem} \label{pr-spectrum}
Show that, for a class of two-body potentials $\Phi$, (\ref{eq-problem1}) holds as the mean-field limit is approached.
\end{problem}

\subsubsection*{Relaxation into Coherent States} 

Next, we describe the problem of {\em ``relaxation into coherent states"} motivated by our expectation that coherent bosonic condensates emerge from non-coherent initial states. 

Recall from equation (\ref{eq-CS}) that derivations of (\ref{eq-HE}) require that the initial $N$-particle wave function, $\Psi_N(0)$, is coherent. Obviously, such initial conditions are exceptional. Therefore it is of fundamental interest to understand how a state, $\Psi_N(t)$, arising from an arbitrary initial condition, $\Psi_N(0)$, approaches (in a sense to be made precise) a coherent state in the limit
\begin{equation} \label{eq-limit}
 t \rightarrow \infty \quad \mbox{and} \quad  N \rightarrow \infty \, , \qquad \nu = \kappa \cdot  N = \mbox{const.} \, ,
\end{equation}
where $t$ will depend on $N$. Thus, we are led to the following problem.

\begin{problem} \label{pr-relax}
Show that, for a class of two-body potentials and a suitable limiting process (\ref{eq-limit}), non-coherent states relax into coherent states. 
\end{problem}

\noindent
{\em Remark.} The study of Problem \ref{pr-relax} necessitates a mean-field spectral analysis of $H_N$ that also takes into account resonances and their life-times. \\

We sketch two different situations, depending on the external potential, and their conjectural qualitative behavior in the limit (\ref{eq-limit}).\\

\begin{figure}[h]
\begin{center}
\psfrag{Psi}{$\Psi_N(t=0)$}
\psfrag{t go infin}{$t \longrightarrow \infty$}
\psfrag{N}{$N$}
\psfrag{N1}{$N_1$}
\psfrag{V}{$\lambda V$}
 \includegraphics[scale=0.7]{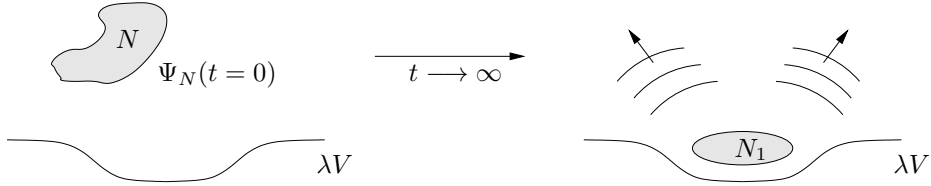}
\end{center}
\caption{\label{fig-trap} ``Trapping of a Hartree Soliton". The $N$-particle state, $\Psi_N(t)$, approaches a coherent state corresponding to a Hartree ground state, $Q$, which is located near the minimum of the trap. Dispersive ``radiation" is emitted during this relaxation process.}
\end{figure}

{\em Trapping.} For localized traps, $\lambda V$, physical intuition suggests a relaxation process corresponding to the ``trapping of a Hartree soliton''. The $N$-particle state, $\Psi_N(t)$, is expected to approach (with positive probability $\varrho > 0$) a coherent bound cluster of bosons described by a Hartree ground state soliton, plus some dispersive ``radiation", in the limit (\ref{eq-limit}). See Figure \ref{fig-trap} for a sketch of this situation. To arrive at a formal statement for this relaxation process, we use the notation of Section \ref{subsec-rigorous}. We expect that
\begin{equation} 
\langle \Psi_N(t), \mathcal{A}_m \Psi_N(t) \rangle \stackrel{(\ref{eq-limit})}{\simeq} \varrho \int_{\RR^{3m} \times \RR^{3m}} a(x_1,\ldots,x_m;y_1, \ldots, y_m) \prod_{j=1}^m Q(x_j) Q(y_j) \, \dd x \, \dd y \, ,
\end{equation}
where $0 \leq \varrho \leq 1$, and $Q$ is a Hartree ground state with $\| Q \|_2^2 = 1$, and where $\nu$ is replaced by $\frac{N_1}{N} \nu > \nu_*$ in (\ref{eq-HE}), with $N_1 < N$. See Section \ref{sec-solitons} for details on Hartree ground states.

In addition to such trapping processes, another situation can occur: If $\Psi_N(t)$ ``scatters'' off the trap, then the $N$-particle state can exhibit the same behavior as described next. \\

\begin{figure}[h]
\begin{center}
\psfrag{v1}{$v_1$}
\psfrag{v2}{$v_2$}
\psfrag{vk}{$v_k$}
\psfrag{N}{$N$}
\psfrag{Psi}{$\Psi_N(t=0)$}
\psfrag{t go infin}{$t \longrightarrow \infty$}
\psfrag{M1}{$N_1$}
\psfrag{M2}{$N_2$}
\psfrag{Mk}{$N_k$}
 \includegraphics[scale=0.7]{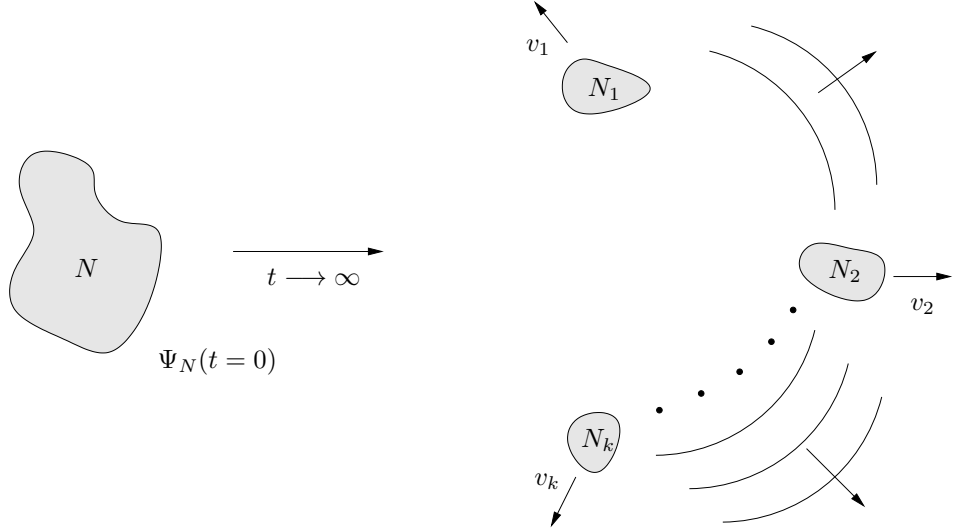}
\end{center}
\caption{\label{fig-hartree} An $N$-particle state, $\Psi_N(t)$, evolving, as $t \rightarrow \infty$, into $k \geq 1$ bound coherent clusters of bosons with $\frac{N_i}{N} \nu > \nu_*$, $i=1, \ldots,k$, receding from each other with velocities, $v_i$, plus some dispersive ``radiation".}
\end{figure}

{\em Vanishing external potential.} Assume that $\lambda=0$, \ie, no external potential is present. If the two-body potential is sufficiently attractive (\ie, $\nu > \nu_*$), then we expect $\Psi_N(t)$ to evolve into bound coherent clusters of bosons receding from each other, plus some small dispersive ``radiation", as $t \rightarrow \infty$; see Figure \ref{fig-hartree}. Using the notation of Section \ref{subsec-rigorous}, the situation envisaged here can be described as follows. Given an $m$-particle operator, $\mathcal{A}_m$, we define $\mathcal{A}_m(Y)$ via its integral kernel
\begin{equation}
a(x_1 - Y, \ldots, x_m-Y; y_1 - Y, \ldots, y_m -Y ) \, , \qquad \mbox{for} \quad Y \in \RR^3 \, .
\end{equation}
In view of (\ref{eq-AExp}), we expect that
\begin{equation}  \begin{split}
\int_{\RR^3} \langle \Psi_N(t), \mathcal{A}_m(Y) \Psi_N(t) \rangle \, \dd Y  \stackrel{(\ref{eq-limit})}{\simeq} & \sum_k \varrho_k \sum_{i=1}^k \int_{\RR^3} \dd Y \int_{\RR^{3m} \times \RR^{3m}}  a(x_1-Y,\ldots,x_m-Y;\\ & y_1-Y,\ldots,y_m-Y)   \prod_{j=1}^m \overline{\psi^{(i)}_k(t,x_j)} \psi^{(i)}_k (t,y_j)  \, \dd x \, \dd y \, . \end{split}
\end{equation} 
Here $\psi^{(i)}_k(t)$ solves (\ref{eq-HE}), with $\| \psi^{(i)}_k(t) \|_2 = 1$ and $\nu_i = \frac{N_i}{N} \nu > \nu_*$, and $\varrho_k$ are non-negative weights with $\sum_k \varrho_k \leq 1$. By physical reasoning, we expect ``ground state selection", \ie, the one-particle wave functions, $\psi^{(i)}_k(t)$,  asymptotically approach {\em Hartree ground states}.

\section{Hamiltonian Structure} \label{sec-hamilton}

Rephrasing partial differential equations as Hamiltonian equations of motion on some infinite-dimensional phase space is a well-known procedure. For the Hartree equation, this also clarifies its {\em point-particle (Newtonian) limit}, which is presented in Section \ref{sec-pplimit} below. 

\subsection{Basic Setup}

We identify the {\em phase space} for (\ref{eq-HE}) with its energy space, \ie, we put   
\begin{equation}
 \Gamma := X = \left \{ \psi \in H^1(\RR^3) : V |\psi|^2 \in L^1(\RR^3) \right \} \, ,
\end{equation}
where it is convenient to treat $\psi$ and $\bar{\psi}$ as canonically conjugate, independent variables. Defining the {\em Hartree Hamilton functional} by
\begin{equation} \label{eq-H} \boxed{
 \mathcal{H}[\bar{\psi}, \psi] := \frac{1}{2} \int_{\RR^3} \left [ | \nabla \psi |^2 + \lambda V |\psi|^2 + \frac{\nu}{2} \big ( \Phi \ast |\psi|^2 \big ) |\psi|^2 \right ] \, \dd x \, ,}
\end{equation}
we are led to the Hartree equation as a Hamiltonian equation of motion:
\begin{equation} \label{eq-HEM}
\dot{\psi}(t) = -i \frac{\delta}{\delta \bar{\psi}} \mathcal{H}[\bar{\psi}(t), \psi(t) ] \; ,
\end{equation}
where $\delta / \delta \bar{\psi}$ denotes the Fr\'echet derivative with respect to $\bar{\psi}$. Alternatively, (\ref{eq-HEM}) can also be expressed in {\em Liouville form}
\begin{equation}
 \dot{\psi}(t) = \big \{ \mathcal{H}[\bar{\psi}(t), \psi(t)], \psi(t) \big \}_{\mathrm{PB}} \, ,
\end{equation}
with {\em Poisson brackets} defined by
\begin{equation}
\{ \psi(x), \psi(y) \}_{\mathrm{PB}} = \{ \bar{\psi}(x), \bar{\psi}(y) \}_{\mathrm{PB}} = 0 \, , \qquad \{ \psi(x), \bar{\psi}(y) \} _{\mathrm{PB}}= 2i \delta(x-y) \, .
\end{equation}
We remark that if $V$ and $\Phi$ meet the conditions of Theorem \ref{th-gwp}, then it is straightforward to show (using Sobolev's inequalities) that $\mathcal{H}$ is a $C^1$-functional on $\Gamma$ bounded from below on the spheres
\begin{equation}
\mathbb{S}_N := \left \{ \psi \in \Gamma : \mathcal{N}[\bar{\psi}, \psi] = N \right \} \, ,
\end{equation}
for all $N < \infty$, where we define the {\em charge} (or {\em mass}) {\em functional} by
\begin{equation} \label{eq-N} \boxed{
\mathcal{N}[\bar{\psi}, \psi ] := \int_{\RR^3} \bar{\psi} \psi \, \dd x \, . }
\end{equation}

Hamilton's equations of motion (\ref{eq-HEM}) can also be derived from an {\em action principle}. The corresponding {\em action functional} is given by
\begin{equation} \label{eq-action}
\mathcal{S}[\bar{\psi}(\cdot), \psi(\cdot) ] := \int_{t_1}^{t_2} \left [ \left ( \frac{i}{2} \int_{\RR^3}  \bar{\psi}(t)  \dot{\psi}(t) \, \dd x \right ) - \mathcal{H}[\bar{\psi}(t), \psi(t) ] \right ] \, \dd t \, ,
\end{equation}
and is defined for all $C^1$-curves $[t_1, t_2] \mapsto \psi(t) \in \Gamma$.  The Hartree equation is equivalent to the Euler-Lagrange equation for $\mathcal{S}[\bar{\psi}, \psi]$ varied with respect to $\bar{\psi}$, \ie,
\begin{equation} \label{eq-EL}
\frac{\delta}{\delta \bar{\psi}(\cdot)} \mathcal{S}[\bar{\psi}(\cdot), \psi(\cdot)]   = 0 \, ,
\end{equation} 
subject to the boundary conditions 
\begin{equation}
\delta \psi(t) |_{t=t_j} = 0 \, , \qquad \mbox{with} \quad j=1,2 \, . 
\end{equation}
Note that the canonically conjugate variables, $\psi$ and $\bar{\psi}$, appear in the action principle, \ie, we consider a Hamiltonian action principle on phase space.

\subsection{Conservation Laws and Identities}

From now on, we abbreviate our notation and write
\begin{equation}
\mathcal{H}[\psi] \equiv \mathcal{H}[\bar{\psi}, \psi] \, , \quad \mathcal{N}[\psi] \equiv \mathcal{N}[\bar{\psi}, \psi] \, , \quad \mathcal{S}[\psi] \equiv \mathcal{S}[\bar{\psi}(\cdot), \psi(\cdot)] \, . 
\end{equation}

\subsubsection*{Charge and Energy} 

The invariance of the action, $\mathcal{S}[\psi]$, with respect to $U(1)$ gauge-transformations and time translations leads to the {\em conservation of charge and energy}, respectively. We have that
\begin{equation}
\mathcal{N}[\psi(t)] = \mathcal{N}[\psi_0] \, , \qquad \mathcal{H}[\psi(t)] = \mathcal{H}[\psi_0] \, ,
\end{equation}
where $\psi(t)$ solves (\ref{eq-HE}). In the setting of Theorem \ref{th-gwp}, these conservation laws hold rigorously and play a central r\^ole in its proof.

\subsubsection*{Linear and Angular Momentum} 

If $\lambda V \equiv 0$ then further conservation laws derived from the Galilean invariance of the action $\mathcal{S}[\psi]$ arise. Because of translation invariance we have {\em conservation of linear momentum:}
\begin{equation} \label{eq-P}
\mathcal{P}[\psi(t)] = \mathcal{P}[\psi_0] \, , \qquad \mbox{where} \quad \mathcal{P}[\psi] := -i \int_{\RR^3} \bar{\psi} \nabla \psi \, \dd x \, .
\end{equation}
Moreover, the {\em center-of-mass motion} is inertial, which is expressed by the fact
\begin{equation} \label{eq-COM}
\int_{\RR^3} \bar{\psi}(t) ( x + 2it \nabla ) \psi(t) \, \dd x = \mathrm{const.}
\end{equation}
Note that, for (\ref{eq-COM}) to hold rigorously, one has to require suitable decay of the initial datum, as $|x| \rightarrow \infty$.  

Finally, from the invariance of $\mathcal{S}[\psi]$ with respect to spatial rotations we can infer {\em conservation of angular momentum:}
\begin{equation}
\mathcal{L}[\psi(t)] = \mathcal{L}[\psi_0] \, , \qquad \mbox{where} \quad \mathcal{L}[\psi] := -i \int_{\RR^3} \bar{\psi} ( x \times \nabla ) \psi \, \dd x \, .
\end{equation}

\subsubsection*{Various Identities}

We note that there are useful ``identities" for nonlinear Schr\"odinger equations, which are not conservation laws. In particular, the {\em pseudo-conformal identity} and {\em Morawetz' identity} are important in scattering theory for repulsive $\Phi$'s. Another identity referred to as {\em variance identity} is an ordinary differential equation. It can be applied to exhibit blow-up of solutions of (\ref{eq-HE}) for non-subcritical, attractive two-body potentials. For example, choosing $\lambda = 0$  and $\nu \Phi(x) = -|x|^{-\sigma}$  the variance identity becomes
\begin{equation} \label{eq-VI}
\frac{\dd^2}{\dd t^2} \int_{\RR^3} |x|^2 |\psi(t)|^2 \, \dd x = 16 \mathcal{H}[\psi_0] + (4-2\sigma) \int_{\RR^3} \big ( |x|^{-\sigma} \ast |\psi(t)|^2 \big ) | \psi(t)|^2 \, \dd x \, ,
\end{equation}  
provided that $\psi_0 \in \left \{ \psi \in X : |x|^2 |\psi|^2 \in L^1(\RR^3) \right \}$. Since the left hand side is the second derivative of a nonnegative function which cannot be strictly concave for all $t \geq 0$, one can infer that $\psi(t)$ can only exist over a finite time interval if the right hand side is bounded from above by a negative constant. This happens for $\sigma \geq 2$ (\ie, $\Phi$ is non-subcritical) and initial data with $\mathcal{H}[\psi_0] < 0$. We refer to \cite{cazenave-03} for a detailed exposition of such identities and their applications to nonlinear Schr\"odinger equations.

\section{Hartree Solitons} \label{sec-solitons}

In this section we discuss solitary wave solutions for (\ref{eq-HE}), which we call {\em Hartree solitons}. By using variational methods, we derive {\em existence and orbital stability} of Hartree ground state solitons, for attractive, subcritical $\Phi$'s. See Theorems \ref{th-existence} and \ref{th-stability} below. Orbital stability for local, subcritical nonlinear Schr\"odinger equations has only been established for special classes, \eg, for pure power nonlinearities and some variations thereof. We also demonstrate how linear stability can be understood from the variational proof of stability. Finally, we touch upon the issue of {\em asymptotic stability}, which involves a detailed study of dispersive estimates of the linearization around a Hartree soliton. Some open problems are proposed.

Plugging the ansatz $\psi(t,x) = e^{i \omega t}Q(x)$, with $\omega \in \RR$, into (\ref{eq-HE}) leads to the nonlinear eigenvalue problem:
\begin{equation} \label{eq-SOL} \boxed{
 \begin{array}{c} \displaystyle -\DD Q + \lambda V Q + \nu \big ( \Phi \ast |Q|^2 \big ) Q = -\omega Q \, , \\ 
 Q \in X, \quad \mathcal{N}[Q] = N \, . \end{array}  }
 \end{equation}
Recall that $X$ denotes the energy space introduced in (\ref{eq-X}). We exclude the trivial solution, $Q \equiv 0$, by requiring that $\mathcal{N}[Q] \equiv \| Q \|_2^2 = N > 0$.

We note that there are two ways of treating problem (\ref{eq-SOL}): Either we fix $N$ and try to solve for $\omega = \omega_N$ and $Q = Q_N$, or we fix $\omega$ and try to solve for $N=N_\omega$ and $Q = Q_\omega$. In physical applications it is sometimes convenient to rephrase the first way by fixing $N$ as an overall constant and then look for solutions $Q=Q_\nu$ and $\omega = \omega_\nu$ depending on the coupling constant $\nu$. 

If there exists a solution $Q=Q_\omega$ of (\ref{eq-SOL}), with $\omega \in I \subset \RR$, then 
\begin{equation}
\psi_\omega (t,x) := e^{i \omega t} Q_\omega(x), \quad \omega \in I \, ,
\end{equation} 
is a family of solitary wave solutions for (\ref{eq-HE}), which we refer to as {\em Hartree solitons}. Provided that $\lambda = 0$ holds, the family of Hartree solitons comprises the functions
\begin{equation} \label{eq-solwav} 
\begin{array}{c} \displaystyle \psi_\sigma (t,x) := e^{i  (  \frac{1}{2} v \cdot (x-d) - \frac{1}{4} v^2 t + \gamma + \omega t  ) } Q_\omega(x - vt - d) \, , \\
\displaystyle \sigma := ( d,v,\gamma,\omega ) \in \RR^3 \times \RR^3 \times [0, 2\pi) \times I \, , \end{array}  
\end{equation}
obtained by applying Galilei transformations to $\psi_\omega(t,x)$; spatial rotations can be omitted if $Q$ is a radial function. See Theorem \ref{th-existence} below for sufficient conditions for $Q$ to be radial.

\subsection{Existence of Hartree Ground States}

We relate the nonlinear eigenvalue problem (\ref{eq-SOL}) to the following variational problem 
\begin{equation}
E(N) := \inf \left \{ \mathcal{H}[\psi] : \psi \in \mathbb{S}_N \right \} \, , 
\end{equation}
where
\begin{equation}
\mathbb{S}_N = \left \{ \psi \in X : \mathcal{N}[\psi] = N \right \} \, .
\end{equation}
If a minimizer exists then it is a solution of (\ref{eq-SOL}), provided that $\mathcal{H}$ is $C^1$. Such energy minimizing solutions are called {\em Hartree ground states}. To arrive at a general statement on the existence of Hartree ground states, we require the following definition: We say that $\Phi$ is {\em subcritical and vanishing at infinity} if
\begin{equation} \label{eq-VAI}
\Phi \in L^{p_1}(\RR^3) + L^{p_2}(\RR^3) \, , \qquad \mbox{for some} \quad 3/2 < p_1 \leq p_2 < \infty  \, ,
\end{equation}
and likewise for $V$.  

\begin{theorem} \label{th-existence} {\bf [Existence of Hartree Ground States]} 
Without loss of generality we set $\nu = 1$ in (\ref{eq-SOL}). Suppose that $\Phi$ is real-valued, radial, subcritical and vanishing at infinity, with $\Phi \leq 0$, $\Phi \not \equiv 0$. Let $V$ be real-valued and assume that one of the following assumptions is satisfied.
\begin{enumerate}
\item[a)] $\lambda = 0$ (\ie, no external potential).
\item[b)] $V$ is subcritical and vanishing at infinity with $V < 0$ on a set of positive measure (\ie, $V$ is a ``localized trap''), $\lambda = 1$.
\item[c)] $V \in L^1_\mathrm{loc}(\RR^3)$, $V \geq 0$, and $ V(x) \rightarrow \infty$ as $|x| \rightarrow \infty$  (\ie, $V$ is a ``confining trap''), $\lambda = 1$. 
\end{enumerate}
Then there exists a constant $N_* \geq 0$ such that, for every $N > N_*$, problem (\ref{eq-SOL}) has a Hartree ground state solution, $Q=Q_N$, which can be chosen to be strictly positive.

In addition, if $\Phi$ is nondecreasing then $Q$ is radial (about some point) and nonincreasing, provided that either a) holds, or b) holds with $V$ radial and nondecreasing.

\end{theorem}

\noindent
{\em Remarks.} 1) Under condition a) one can show that $N_*=0$, for $\Phi(x) = -|x|^{-\sigma}$, with $0 < \sigma < 2$. However, for two-body potentials of short range (\eg, of van der Waals-type) $N_*$ is strictly positive. With regard to condition c), it is easy to see that $N_*=0$, for confining $V$'s.

2) By applying standard methods we can derive decay and regularity properties for $Q_N$ depending on $V$. If, for instance, condition a) holds we can infer that $Q_N \in C^\infty(\RR^3)$, thanks to elliptic bootstrapping, and that $Q_N$ exhibits exponential decay at infinity.

3) Theorem \ref{th-existence} can be rephrased as follows: For $N \equiv 1$ there exists a critical coupling, $\nu_* \geq 0$, such that, for all $\nu > \nu_*$, there exists a Hartree ground state $Q =Q_\nu$. \\

\noindent
{\em Sketch of Proof.} The assumptions of Theorem \ref{th-existence} imply that, for every $ \varepsilon \in (0,1)$, there exists a constant $C_{\varepsilon,N} \geq 0$ such that
\begin{equation} E(N) \geq \frac{1-\varepsilon}{2} \| \nabla \psi \|_2^2 - C_{\varepsilon, N} \, . \end{equation}
In particular, this shows that $E(N) > -\infty$ holds for all $N < \infty$. To prove existence of Hartree ground states, for $N$ sufficiently large, we apply the direct method of the calculus of variations. That is, we consider minimizing sequences, $(\psi_n)_{n \geq 0}$, in $\mathbb{S}_N$ with
\[
\mathcal{H}[\psi_n] \rightarrow E(N) \, , \qquad \mbox{as} \quad n \rightarrow \infty  \, ,
\]
and try to find a strongly convergent subsequence in $X$. Depending on conditions a), b), or c) in Theorem \ref{th-existence}, we can show relative compactness of every minimizing sequence as follows.

{\em Relative Compactness under Condition a).} Due to the assumptions on $\Phi$, we can apply the concentration-compactness method developed by P.-L.~Lions \cite{lions-84a,lions-84b} for translation-invariant functionals. In these references, it is shown that the {\em strict sub-additivity condition} 
\begin{equation} \label{eq-SAC} 
 E(N) < E(\alpha) + E(N-\alpha) \, , \qquad \mbox{for all} \quad \alpha \in (0, N )  \, ,
\end{equation}
is sufficient and necessary to show that every minimizing sequence, $(\psi_n)_{n \geq 0}$, of $\mathcal{H}[\psi] |_{\psi \in \mathbb{S}_N}$ is relatively compact in $X \equiv H^1(\RR^3)$ up to translations. Thus, there exists a sequence, $(y_n)_{n \geq 0}$, in $\RR^3$ such that $(\psi_n(\cdot + y_n))_{n \geq 0}$ has a subsequence that is strongly convergent in $H^1(\RR^3)$. In particular, a minimizer exists if this property holds. 

We must show that (\ref{eq-SAC}) holds for all $N > N_*$, for some constant $N_* \geq 0$. This is accomplished as follows. First, it is easy to see that
\begin{equation}
 E(N) \leq 0 \, , \qquad \mbox{for all} \quad N \geq 0 \,  . 
\end{equation}
Moreover, we infer from the scaling behavior
\begin{equation}
E(\vartheta N ) = \inf \Big \{ \frac{\vartheta}{2}\int_{\RR^3} \big [ | \nabla \psi |^2 + \frac{\vartheta}{2} \big ( \Phi \ast |\psi|^2 \big ) | \psi |^2 \big ] \, \dd x : \psi \in \mathbb{S}_N \Big \} 
\end{equation}
that there exists some $N_* \geq 0$ such that 
\begin{equation}
 E(N) < 0 \,  , \qquad \mbox{for all} \quad  N > N_* \, , 
 \end{equation}
and that
\begin{equation}
E(\vartheta N) < \vartheta E(N) \, , \qquad  \mbox{for all} \quad  N > N_* \quad \mbox{and} \quad \vartheta > 1\, .
\end{equation}  
The strict sub-additivity condition (\ref{eq-SAC}) for $N > N_*$ then follows by elementary arguments.

{\em Relative Compactness under Condition b).} To show that every minimizing sequence for $N > N_*$ has a strongly convergent subsequence in $X \equiv H^1(\RR^3)$ (up to translations), we can proceed similarly as in a). See \cite{lions-84a,lions-84b} for a modified sub-additivity condition for translation-dependent functionals.

{\em Relative Compactness under Condition c).} If $V$ is confining then it easy to see that, for every $\varepsilon > 0$, there exist constants $R_\varepsilon, n_\varepsilon  \geq 0$ such that
\[
\int_{ |x| \geq R_\varepsilon} |\psi_n|^2 \, \dd x \leq \varepsilon \, , \qquad \mbox{for all} \quad n \geq n_\varepsilon \, .
\]
This fact allows us to pass to a strongly convergent subsequence for every minimizing sequence in $X$.

{\em Strict Positivity.} Note that $\mathcal{H}[|\psi|] \leq \mathcal{H}[\psi]$, because $\| \nabla |\psi| \|_2 \leq \| \nabla \psi \|_2$. Thus, we can choose $Q_N$ to be real-valued and nonnegative. Introducing the Schr\"odinger operator
\begin{equation} 
 H := -\DD + W \, , \qquad \mbox{where} \quad W:= V + \big ( \Phi \ast |Q_N|^2 \big ) \, ,
\end{equation}
we have that $Q_N$ is an eigenfunction, \ie,
\begin{equation} H Q_N = -\omega_N Q_N \, , \end{equation}
which can be written as follows
\begin{equation}
  e^{-H} Q_N = e^{\omega_N} Q_N \, . 
\end{equation}
The assumptions of Theorem \ref{th-existence} imply that $W_+ := \max \{ W, 0 \} \in L^1_{\mathrm{loc}}(\RR^3)$ and that $W_- := \min \{ W, 0 \}$ is subcritical. With results from \cite{farissimon-74} we find that $e^{-H}$ is {\em positivity improving}, \ie, $f \geq 0$, with $f \not \equiv 0$, implies that $e^{-H} f> 0$. Hence $Q_N(x) > 0$ almost everywhere.

{\em Radial Symmetry.} The statement about radial symmetry of $Q_N$ in Theorem \ref{th-existence} follows from
\begin{equation}
\mathcal{H}[Q_N^*] \leq \mathcal{H}[Q_N] \, ,
\end{equation}
provided that $\Phi = \Phi^*$ and $V = V^*$, where $f^*$ denotes the symmetric rearrangement of a measurable function $f$ that vanishes at infinity; see, \eg, \cite{liebloss-01} for rearrangement inequalities. $\qed$ \\

\noindent 
{\em Remarks.} 1) In physical terms, condition (\ref{eq-SAC}) says that Hartree ground states are stable against fission: A ground state configuration having mass $N$ has an energy strictly smaller than the sum of energies of two configurations with mass $\alpha$ and $N-\alpha$, respectively. Hence there is always some strictly negative binding energy. In mathematical terms, condition (\ref{eq-SAC}) prevents every minimizing sequence from either converging to zero locally everywhere, or splitting up into lumps receding from each other.

2) If $N$ is sufficiently large then nonlinear excited states (which are not strictly positive) exist, too. They are critical points of $\mathcal{H}[\psi]$ restricted to $\mathbb{S}_N$; see \cite{lions-80} for a critical point theory. \\  

Theorem \ref{th-existence} does not answer the question of {\em uniqueness} of Hartree ground states, \ie, whether all minimizers of $\mathcal{H}[\psi]$ on $\mathbb{S}_N$ are just the orbit of some $Q_N$ under the group of symmetries leaving $\mathcal{H}[\psi]$ invariant. By using a remarkable trick, E.~Lieb \cite{lieb-77} has proven uniqueness of Hartree minimizers for (\ref{eq-bsnr}), meaning that all minimizers are equivalent up to phases and translations. But the general case is not understood, yet. Thus, we propose.

\begin{problem} \label{pr-uniqueness}
Prove uniqueness of Hartree ground states on $\mathbb{S}_N$, up to symmetries of $\mathcal{H}[\psi]$, for a general class of $\Phi$'s and $\lambda = 0$. 
\end{problem} 
 
 \noindent
 {\em Remarks.} 1) Uniqueness of positive ground state solutions for nonlinear Schr\"odinger equations on $\RR^d$ with local nonlinearities of the form $|\psi|^{\alpha} \psi$ is a known fact, for $0 < \alpha < 4/(d-2)$; see \cite{kwong-89}. The proof given there relies, however, on Sturm comparison theorems and ordinary differential equation techniques that cannot be directly applied to nonlocal equations such as (\ref{eq-SOL}). 
 
2) If an external potential is present (\ie, $\lambda \neq 0$), then one can also study perturbative regimes to show existence and uniqueness of Hartree ground states. Assuming that $-\DD + \lambda V$ has a normalizable ground state, one can prove uniqueness of Hartree ground states for $\mathcal{H}[\psi] |_{\mathbb{S}_N}$, with $N$ fixed and $\nu$ sufficiently small; see, \eg, \cite{afgst-02} for such results.\\

Finally, we wish to draw attention to the phenomenon of {\em symmetry breaking:} If, for example, $V$ is a double-well potential then, for suitable $\Phi$ and sufficiently large $\nu$, two distinct Hartree ground states exist; \ie, uniqueness is lost \cite{afgst-02}. (Of course, translation invariance is always broken when $\lambda = 0$, as soon as a non-trivial Hartree ground state exists.)

\subsection{Orbital Stability of Hartree Ground State Solitons}

Next, we study orbital stability of Hartree solitons arising from the Hartree ground states constructed in Theorem \ref{th-existence}. For this purpose, we introduce the set of Hartree ground states, $\mathcal{M}_N$, for $N > N_*$, by
\begin{equation}
\mathcal{M}_N := \big \{ Q \in \mathbb{S}_N : \mathcal{H}[Q] = \min_{\psi \in \mathbb{S}_N} \mathcal{H}[\psi] \big \} \, , 
\end{equation}
and define the distance function
\begin{equation}
 d(\psi) := \inf_{Q \in \mathcal{M}_N} \| \psi - Q \|_X \, .
\end{equation}
We note that if we have uniqueness of Hartree ground states (up to phases and translations), then 
\begin{equation} \label{eq-inf2}
 d(\psi) \equiv \inf_{\gamma \in \RR} \inf_{y \in \RR^3} \| \psi - e^{i \gamma} \widetilde{Q}(\cdot - y ) \|_{X} 
\end{equation}
for a fixed $\widetilde{Q} \in \mathcal{M}_N$. Recall that uniqueness of Hartree ground states is known for equation (\ref{eq-bsnr}), which models a non-relativistic boson star; see \cite{lieb-77}.

We say that $\mathcal{M}_N$ is {\em orbitally stable in $X$} (energy space) if, for every $\varepsilon > 0$, there exists $\delta > 0$ such that, for all initial data $\psi_0 \in X$,  
\begin{equation} \label{eq-inf}
 d(\psi_0) < \delta \qquad \mbox{implies that} \qquad  d(\psi(t)) < \varepsilon \,  , \quad \mbox{for all} \quad t \geq 0 \, ,
 \end{equation}
where $\psi(t)$ denotes the solution of (\ref{eq-HE}). With the proof of Theorem \ref{th-existence} at hand, it is easy to prove the following result.

\begin{theorem} \label{th-stability} {\bf [Orbital Stability of Hartree Ground State Solitons]}
Assume that the hypotheses of Theorem \ref{th-existence} are satisfied (and if $V$ is confining that condition b) of Theorem \ref{th-gwp} holds). Then orbital stability of the set of Hartree ground states, $\mathcal{M}_N$, holds for all $N > N_*$. 
\end{theorem}

\noindent
{\em Remark.} If the external potential vanishes (\ie, $\lambda = 0$), then Theorem \ref{th-stability} also guarantees orbital stability of ``boosted'' Hartree ground state solitons; see equation (\ref{eq-solwav}). This fact follows by using the Galilei invariance of (\ref{eq-HE}).\\

\noindent
{\em Sketch of Proof.}  The proof of Theorem \ref{th-stability} makes use of conservation laws and the relative compactness of every minimizing sequence; the idea goes back to \cite{cazlions-82}. It proceeds as follows. Recall from Theorem \ref{th-gwp} that (\ref{eq-HE}) is globally well-posed in $X$. To simplify the argument, we restrict our attention to initial data with $\mathcal{N}[\psi_0] =N$ (the general case will then follow). Let us assume that orbital stability does not hold. Then there would exist an $\varepsilon > 0$ and a sequence $(u_n)_{n \geq 0}$ in $\mathbb{S}_N$ with
\begin{equation} 
d(u_n) \rightarrow 0 \, , \qquad \mbox{as} \quad  n \rightarrow \infty  \, ,\end{equation}
such that, for the solution $\psi_n(t)$ of (\ref{eq-HE}) with initial datum $u_n$, 
\begin{equation} \label{eq-nn}
 d(\psi_n(t_n)) \geq \varepsilon \, , \qquad \mbox{for all} \quad n \geq 0 \, 
\end{equation}
and a suitable sequence $(t_n)_{n \geq 0} $ of times. Let us define $v_n := \psi_n(t_n)$. Since charge and energy are conserved, \ie, 
\begin{equation}
\mathcal{N}[ v_n ] = \mathcal{N}[ u_n ] \, , \qquad  \mathcal{H}[v_n] = \mathcal{H}[u_n] \, ,
\end{equation}
the sequence $(v_n)_{n \geq 0}$ is also a minimizing sequence for $\mathcal{H}[\psi] |_{\mathbb{S}_N}$. As we have already seen in the proof of Theorem \ref{th-existence}, every such minimizing sequence contains a strongly convergent subsequence in $X$ (up to possible translations). This contradicts (\ref{eq-nn}) and proves Theorem \ref{th-stability}.  $\qed$

\subsubsection*{Orbital Stability via Linearization}

There is another way to discuss orbital stability involving the linearization around a Hartree soliton. In fact, a rigorous analysis, initiated by M.~Weinstein \cite{weinstein-85} for local nonlinear Schr\"odinger equations and extended by M.~Grillakis, J.~Shatah, and W.~Strauss \cite{grillakis-90}, shows that the {\em stability condition}
\begin{equation} \label{eq-STAB}
 \frac{\dd}{\dd \omega} \mathcal{N}[Q_\omega] > 0 
\end{equation}
is sufficient for orbital stability of Hartree ground states. Recall that $Q=Q_\omega$ denotes the solution of (\ref{eq-SOL}), for a given $\omega$, choosing $N=\mathcal{N}[Q_\omega]$. Condition (\ref{eq-STAB}) is popular in the physics literature on solitons. 

Under the assumptions of Theorem \ref{th-existence}, we now demonstrate why condition (\ref{eq-STAB}) implies the strict sub-additivity condition (\ref{eq-SAC}) and thus orbital stability. It turns out that the relation between these two criteria is essentially the same as the one between the inequality $f'' < 0$ and the strict concavity of a function $f$. 

To convince ourselves of this claim, we assume that $Q=Q_N$ and $Q=Q_\omega$ can be differentiated with respect to $N$ or $\omega$ up to second order. We observe that the nonlinear eigenvalue problem (\ref{eq-SOL}) with constraint $\mathcal{N}[Q] = N$ is just the Lagrange equation with multiplier $-\omega/2$, \ie, 
\begin{equation} \label{eq-hq}
\mathcal{H}'[Q_N] = - \frac{\omega}{ 2} \mathcal{N}'[Q_N] \, . 
\end{equation}
Thus, for all $N > N_*$, we find that
\begin{equation} \begin{split}
 E'(N)  & =   \frac{\dd}{\dd N} \mathcal{H}[Q_N] =  \mathcal{H}'[Q_N] \frac{\dd Q_N}{\dd N} \stackrel{(\ref{eq-hq})}{=} -\frac{\omega}{2} \mathcal{N}'[Q_N] \frac{\dd Q_N}{\dd N}  \\ & =    -\frac{\omega}{2} \frac{\dd \mathcal{N}[Q_N]}{\dd N} = -\frac{\omega}{2} \frac{\dd N}{\dd N} = -\frac{\omega}{2} \, . \end{split}
\end{equation}
The stability condition (\ref{eq-STAB}) then implies that 
\begin{equation} E''(N) = - \frac{1}{2} \frac{\dd \omega}{ \dd N}  < 0 \, . \end{equation}
Hence $E(N)$ is a strictly concave function for all $N > N_*$. The strict sub-additivity condition (\ref{eq-SAC}) is a direct consequence of this fact.

\subsection{Instability and Blow-Up}

Using the concentration-compactness method to prove orbital stability is quite elegant. It is, however, not suited to disprove orbital stability if $\Phi$ is not subcritical, because it is possible that $E(N) = -\infty$, \ie, that the Hartree energy functional is not bounded from below on $\mathbb{S}_N$ (hence $\mathcal{M}_N$ is empty). If this happens, linearization around Hartree solitons, defined to be critical points of $\mathcal{H}[\psi]$ on $\mathbb{S}_N$, appears to be indispensable. In fact, it can be shown that the {\em instability condition}
\begin{equation} \label{eq-ISTAB}
 \frac{\dd}{\dd \omega} \mathcal{N}[Q_\omega] < 0 
\end{equation}
is sufficient for {\em orbital instability} for $Q = Q_\omega$ \cite{grillakis-90}. This means that there exists an $\varepsilon > 0$, but no $\delta > 0$ such that $d(\psi_0) < \delta$ implies that $d(\psi(t)) < \varepsilon$, for all $t \geq 0$, where $d(\cdot)$ is defined as
\begin{equation}
d(\psi) := \inf_{\gamma \in \RR} \inf_{y \in \RR^3} \| \psi - e^{i\gamma} Q(\cdot - y) \|_X \, .
\end{equation}

As an example, we illustrate instability of Hartree solitons for the non-subcritical case,
\begin{equation} \label{eq-mod}
i \partial_t \psi = -\DD \psi - \big ( |x|^{-\sigma} \ast |\psi|^2 \big ) \psi \, , \qquad \mbox{with} \quad 2 \leq \sigma < 3 \, ,
\end{equation}
where non-trivial solutions, $Q=Q_\omega$, of the equation
\begin{equation} \label{eq-qw}
-\DD Q - \big ( |x|^{-\sigma} \ast |Q|^2 \big ) Q = -\omega Q \, ,
\end{equation}
are known to exist; see \cite{lions-80}. A calculation shows that 
\begin{equation}
\mathcal{N}[Q_\omega] = C \omega^{1-\frac{\sigma}{2}} 
\end{equation}
for some positive constant $C=C(\sigma)$. Hence, the instability condition (\ref{eq-ISTAB}) holds for $2 < \sigma < 3$. That orbital instability also occurs for $\sigma = 2$ (where condition (\ref{eq-ISTAB}) fails) can be demonstrated with the help of the variance identity (\ref{eq-VI}) as follows. Multiplying equation (\ref{eq-qw}) by $\bar{Q}$ and integrating over space, and doing the same with $x \cdot \nabla \bar{Q}$, leads to the identity  
\begin{equation}
\mathcal{H}[Q] = 0 \, .
\end{equation}
Choosing an initial datum $\psi_0 := (1 + \varepsilon) Q$, we find that $\mathcal{H}[\psi_0] < 0$, for all $\varepsilon > 0$.  According to the variance identity (\ref{eq-VI}) and the discussion following it, the corresponding solution of (\ref{eq-HE}) exhibits blow-up in finite time; (note that $\psi_0$ has finite variance because of its exponential decay). Choosing $\varepsilon$ sufficiently small we can approximate $Q$ by $\psi_0$ arbitrarily well in $X \equiv H^1(\RR^3)$. This shows instability of $\psi(t,x)=e^{i \omega t}Q(x)$ in (\ref{eq-mod}), for $\sigma = 2$.

\subsection{Asymptotic Stability}

Refining the stability analysis leads one to the issue of {\em asymptotic stability} of Hartree solitons. One asks whether in some suitable norm, $\| \cdot \|_Y$,
\begin{equation}
\big \| \psi(t,\cdot) - e^{i \vartheta(t,x)} Q_{\omega(t)} (\cdot-y(t)) \big \|_Y \rightarrow 0 \, , \qquad \mbox{as} \quad t \rightarrow +\infty \, ,
\end{equation} 
for all initial data $\psi_0$ sufficiently close to a soliton configuration and some functions $\vartheta(t,x), \omega(t)$, and $y(t)$ with the properties that
\begin{equation}
\vartheta(t,x) \rightarrow \frac{1}{2} v_\infty \cdot (x-y_\infty) - \frac{1}{4} v_\infty^2 t + \gamma_\infty + \omega_\infty t \, , \quad \omega(t) \rightarrow \omega_\infty \, , \quad y(t) \rightarrow y_\infty + v_\infty t  \, ,
\end{equation}
as $t \rightarrow +\infty$. A generalization of the notion of asymptotic stability to multi-soliton configurations can be formulated easily.

We briefly outline the history of work on asymptotic stability for nonlinear Schr\"odinger equations. First results date back to the work of A.~Soffer and M.~Weinstein \cite{sofferweinstein-90} for local nonlinear Schr\"odinger equations with suitable external potentials and further restrictive conditions. Their work has been motivated, in part, by conjectures of J.~F.~and T.~Spencer. (For nonlinear Klein-Gordon equations with spatial inhomogeneities, first results appeared in \cite{si-93} and were extended in \cite{sw-99}). The first work on asymptotic stability for translation-invariant nonlinear Schr\"odinger equations (also under very restrictive conditions) was presented by V.~Buslaev and G.~Perel'man \cite{bp-92}. Both works were significantly extended in \cite{sofferweinstein-92, sofferweinstein-03, bp-95, cuccagna-02, ty-02a, ty-02b, ty-02c, gnt-03}. See \cite{pilletwayne-97, weder-00} for a center-manifold approach. Recent results concerning scattering of multi-soliton configurations for local nonlinear Schr\"odinger equations have been established in \cite{perelman-03,rodnianski-03}; (see also \cite{fyt-02} for partial results).

As far as extending such methods and results to Hartree-type nonlinearities is concerned, one expects that there are natural conditions on $\Phi$ allowing for  asymptotic stability. Such results would have interesting physical applications. 

To propose a problem concerning asymptotic stability of Hartree solitons, we assume that $\lambda = 0$  and consider the linearization of (\ref{eq-HE}) around a Hartree ground state $Q_\omega$. Making the ansatz
\begin{equation}
\psi(t,x) = e^{i \omega t} [ Q_\omega(x) + h(t,x) ]
\end{equation}
leads to the linear equation
\begin{equation}
 i \partial_t h = L_\omega h \, ,
\end{equation}
where
\begin{equation}
L_\omega h = -\DD h + \omega h + \big (\Phi \ast |Q_\omega|^2 \big ) h + 2 Q_\omega \big ( \Phi \ast (Q_\omega ( h + \bar{h} )) \big ) \, .
\end{equation}  
Note that $L_\omega$  is not $\mathbb{C}$-linear and nonlocal. As an immediate consequence of the symmetries of (\ref{eq-HE}) with respect to $U(1)$ gauge-transformations and spatial translations one finds that
\begin{equation}
L_\omega i Q_\omega = 0 \, , \quad L_\omega \partial_j Q_\omega = 0 \, , \quad \mbox{where} \quad j = 1,2,3  \, . 
\end{equation} 
This observation leads to the following problem.
\begin{problem} \label{pr-L}
Find a class of two-body potentials $\Phi$ with the following properties.
\begin{enumerate}
\item[a)] One has dispersive estimates for $L_\omega$.
\item[b)] The null space of $L_\omega$ is given by
\[ \mathrm{Null} \, (L_\omega ) = \mathrm{span}_{\RR} \, \{ i Q_\omega, \partial_j Q_\omega \} \, ,\qquad \mbox{where} \quad j=1,2,3 \, , \]
as expected on the basis of the symmetries of (\ref{eq-HE}).
\end{enumerate}
\end{problem}  

\noindent
{\em Remark.} Note that the local potential, $\Phi \ast |Q_\omega|^2$, in $L_\omega$ exhibits power-law decay if $\Phi$ has power-law decay. In contrast, local nonlinearities lead to exponential decay, thanks to the exponential decay of $Q_\omega$. Exponential decay of the background potential in the linearization around a soliton plays a central r\^ole in the derivation of dispersive estimates in \cite{rodnianski-03}.

\section{Point-Particle Limit} \label{sec-pplimit}

In this section, we discuss the {\em point-particle (Newtonian) limit} of the Hartree equation (\ref{eq-HE}) and show how {\em Newton's equations of motion} for a system of point-particles emerge. 

\subsection{Basic Ideas and Result}    

To exhibit the point-particle limit, we consider the following situation. The external potential is given by
\begin{equation}
 \lambda V(x) := V^{(\varepsilon)} (x) := W(\varepsilon x) \, ,
\end{equation}
where $W(x)$ is a smooth, bounded function and $\varepsilon > 0$ denotes a parameter. Furthermore, we assume the two-body potential to be given by
\begin{equation}
 \Phi(x) = \Phi_s(x) + \Phi_\ell (\varepsilon x) \, .
\end{equation}
Here $\Phi_s$ is a radial, short-range, attractive potential, whereas $\Phi_\ell$ is smooth, radial and may be either attractive or repulsive. In addition, $\Phi_\ell$ is allowed to be of long range, \eg, the Coulomb potential.

As a starting point, let us consider a configuration of Hartree ground states for $\Phi = \Phi_s$ (\ie, $\varepsilon = 0$). We pick $k$ numbers $N_1,\ldots, N_k >  N_*(\Phi_s)$ and Hartree ground states, $Q_{N_j}$, minimizing $\mathcal{H}[\psi]$ on $\mathbb{S}_{N_j}$, for $\lambda = 0$ and $\Phi_\ell \equiv 0$ (see Theorem \ref{th-existence}). We consider an initial configuration of $k$ solitons
\begin{equation}
\left \{ Q_{N_j}(x-q_j) \right \}_{j=1}^k 
\end{equation}
centered at points $q_j \in \RR^3$, $1 \leq j \leq k$. These solitons have a characteristic diameter given by
\begin{equation}
 L_{\mathrm{sol}} := \max_{j =1, \ldots, k }  \sqrt{\frac{1}{ \omega_j}} \, .
\end{equation} 
The external potential, $V^{(\varepsilon)}(x) = W(\varepsilon x)$, defines another characteristic length  
\begin{equation}
L_{\mathrm{ext}} := \frac{1}{ \varepsilon \sup | \nabla W| } \, .
\end{equation}
Choosing $\varepsilon > 0$ sufficiently small, such that
\begin{equation}
 \frac{L_{\mathrm{sol}}}{ L_{\mathrm{ext}}} =  \varepsilon \cdot \frac{ \sup | \nabla W|}{L_{\mathrm{sol}}} \ll 1\, ,
\end{equation}
one finds that the external potential varies little over the typical spatial extension of a soliton, \ie, the solitons are almost ``point-like".  Finally, we adjust the initial positions of the solitons to be far separated from each other, \ie, 
\begin{equation} \label{eq-eps}
 \varepsilon \geq \frac{ L_{\mathrm{sol}} }{ \min_{i \neq j} | q_i - q_j | } \, . 
\end{equation}

Having specified our initial configuration, we now look for solutions of (\ref{eq-HE}) of the form
\begin{equation} \label{eq-ansatz}
\psi^{(\varepsilon)}(t,x) = \sum_{j=1}^k Q_{N_j(t)} (x-q_j(t)) e^{i \theta_j(t,x)} + h^{(\varepsilon)}(t,x) 
\end{equation}
obeying, for all times $t$ with $|t| < \mathcal{O}(\varepsilon^{-1})$, the following equations 
\begin{align}
& \theta_j(t,x) =  m \dot{q}_j(t) \cdot [x-q_j(t)] + \vartheta_j(t) \, , \\
&| \dot{N}_j(t) | = o(\varepsilon) \, , \\
& \|  h^{(\varepsilon)}(t,\cdot)  \|_{H^1}= o(\varepsilon) \, ,
\end{align}
where we reinstall the boson mass $m$. A formal calculation with the action (\ref{eq-action}), using the ansatz (\ref{eq-ansatz}), yields
\begin{equation}
\begin{split}
\mathcal{S} \big [\bar{\psi}^{(\varepsilon)}, \psi^{(\varepsilon)} \big ]   = & \int_{-\tau}^\tau\sum_{j=1}^k \Big[  \frac{m}{2} N_j \dot{q}_j(t)^2 - N_j W (\varepsilon q_j(t))  \\
&  - \sum_{i < j} N_i N_j \Phi_\ell \left ( \varepsilon(q_i(t)-q_j(t)) \right ) + o(\varepsilon) \Big ] \, \dd t  \, ,
\end{split}
\end{equation}
for $\tau = \mathcal{O}(\varepsilon^{-1})$. For the action to be stationary at $\psi = \psi^{(\varepsilon)}$, meaning that $\psi^{(\varepsilon)}$ solves (\ref{eq-HE}), the set of trajectories, $\{ q_j(t) \}_{j=1}^k$, must apparently satisfy {\em Newton's equations of motion} for $k$ point-particles with mutual interactions given by $\Phi_\ell$ and subject to an external potential $W$, up to small friction forces:
\begin{equation} 
m \ddot{q}_j(t) =  - \varepsilon (\nabla W)(\varepsilon q_j(t))  - \varepsilon \sum_{i \neq j} N_i (\nabla \Phi_\ell) ( \varepsilon (q_j(t)-q_i(t))) + a_j(t) \, , 
\end{equation}
for $j = 1, \ldots, k$ and times $t$ with  $|t| < \mathcal{O}(\varepsilon^{-1})$. The inhomogeneities, $a_j(t)$, obey the estimate
\begin{equation}
|a_j(t)| = o(\varepsilon) \, ,
\end{equation} 
and are interpreted as ``radiation damping" or ``friction".

All these considerations can be formulated as a theorem for $k=1$ \cite{fyt-02}, but the methods can be extended to general $k$. A more systematic approach to such problems can be found in \cite{fgjs-04}, where use of the Hamiltonian formalism and the associated symplectic structure is made.    

\section{More Open Problems, Conclusions, and Outlook} \label{sec-end}

Recall the open problems already mentioned above: Problem \ref{pr-spectrum} (mean-field spectral analysis),  Problem \ref{pr-relax} (relaxation into coherent states), Problem \ref{pr-uniqueness} (uniqueness of Hartree ground states), and Problem \ref{pr-L} (dispersive estimates for asymptotic stability). In addition to these problems, we wish to draw the reader's attention to the following open problems, which we think are worth studying.

\subsubsection*{Integrable Motions} 
Consider (\ref{eq-HE}) with an external harmonic trap:
\begin{equation}  \label{eq-KAM}
 i \partial_t \psi = -  \DD \psi +  |x|^2 \psi + \big ( \Phi \ast |\psi|^2 \big ) \psi \, .
\end{equation}
One easily verifies that if $Q(t,x)$ solves (\ref{eq-KAM}) then another solution is given by
\begin{equation}
  \psi(t,x) := e^{i[p_\mathrm{c}(t) \cdot x - Et]} Q(t, x-x_\mathrm{c}(t)) \, ,
\end{equation}
where $\{ p_\mathrm{c}(t), x_\mathrm{c}(t) \}$ solve the classical equations of motion of a harmonic oscillator in $\RR^3$, \ie,
\begin{equation}
\dot{p}_\mathrm{c}(t) = -2 x_\mathrm{c}(t) \, , \quad \dot{x}_\mathrm{c}(t) = 2 p_\mathrm{c}(t) \, , \qquad \mbox{where} \quad  E := (p_\mathrm{c}(t))^2 + (x_\mathrm{c}(t))^2 = \mbox{const}. \, 
\end{equation}
Thus, we can construct solutions of (\ref{eq-KAM}) corresponding to Hartree solitons ``moving'' along orbits of an integrable point-particle system. 

The following generalization is of interest: Consider an external potential $V$ that is radially symmetric. Then the Hamilton function
\begin{equation} \label{eq-KAM2}
p^2 + V(x)
\end{equation}
describes an integrable Hamiltonian system. Choose $V$ such that the system satisfies strong anisochronicity conditions. Using ``singular'' KAM theory, attempt to construct integrable motions of Hartree solitons, for $N$ sufficiently large, whose centers of mass follow orbits of (\ref{eq-KAM2}).

\subsubsection*{Formation of Hartree Solitons} 

In view of physical applications, \eg, to boson stars, it would be interesting to gain a deeper understanding of the formation of Hartree solitons. Consider, for example, a configuration of two far-separated solitons with masses $N_1, N_2$, initial positions $q_1,q_2$, and initial velocities $v_1 = \dot{q}_1(0), v_2= \dot{q}(0)$, as described in equation (\ref{eq-ansatz}), with $h^{(\varepsilon)}(x,0)=0$. Assuming that the two-body potential, $\Phi_\ell$, is purely attractive and of short range, we choose $ \{ N_i, q_i, v_i \}_{i=1}^2$ in such a way that\begin{equation}
N_1 v_1^2 + N_2 v_2^2 + N_1 N_2 \Phi_\ell( \varepsilon(q_1 - q_2) ) < 0 \, ,
\end{equation}
\ie, we consider a {\em bound} configuration of two solitons. What happens to this system as $t \rightarrow \infty$? Intuitively, we expect collapse into a single soliton moving inertially through space (for $\lambda = 0$), perturbed by some outgoing dispersive radiation tending to zero at the free dispersion rate. There is numerical evidence for such a behavior, but analytical results are still lacking.

We also mention recent results on soliton formation derived in \cite{tao-04}. The ideas developed there appear applicable to (\ref{eq-HE}) with attractive two-body potentials of short range.

\subsubsection*{Random External Potentials}

In the study of transport phenomena, it is interesting to consider (\ref{eq-HE}) with a {\em random} external potential, $V$, whose distribution is homogeneous and has correlations of short range; see \cite{af-88,afs-88,bw-04} for local nonlinear Schr\"odinger equations with random external potentials, and \cite{fsw-86} for KAM methods applicable to some random systems with infinitely many degrees of freedom.

\subsubsection*{Renormalization Group Methods}

Recent work on gravitational collapse incorporates ideas from the theory of critical phenomena and renormalization group methods; see, \eg, \cite{gundlach-99}. It would be of interest to apply such methods to the semi-relativistic Hartree equations (\ref{eq-bs}) modeling a boson star; see \cite{kup-95} for a general discussion of renormalization group methods in the analysis of blow-up of solutions of nonlinear partial differential equations.

\bigskip

\centerline{---------}

\bigskip

{\small

\bibliographystyle{amsalpha}
\bibliography{QuantumBoseHartreeBib}
}

\vspace{1cm}

\noindent 
{\sc J\"urg Fr\"ohlich\\
Institute for Theoretical Physics\\
ETH Z\"urich-H\"onggerberg\\
CH-8093 Z\"urich, Switzerland}\\
{\em E-mail address:} {\tt juerg@itp.phys.ethz.ch}
 
\vspace{1cm}

\noindent
{\sc Enno Lenzmann\\
Department of Mathematics\\ 
ETH Z\"urich\\
CH-8092 Z\"urich, Switzerland}\\
{\em E-mail address:} {\tt lenzmann@math.ethz.ch}

\end{document}